\documentclass{elsart}
\usepackage{graphicx,harvard,amssymb}

\makeatletter
\def\elsartstyle{%
	\def\normalsize{\@setfontsize\normalsize\@xiipt{14.5}}
	\def\small{\@setfontsize\small\@xipt{13.6}}
	\let\footnotesize=\small
	\def\large{\@setfontsize\large\@xivpt{18}}
	\def\Large{\@setfontsize\Large\@xviipt{22}}
	\skip\@mpfootins = 18\p@ \@plus 2\p@
	\normalsize
}
\makeatother

\def\bibcode#1{}

\def\url#1{{\ttfamily\def\/{/\discretionary{}{}{}}#1}}

\begin{document}

\begin{frontmatter}
\title{The Polarisation Signatures of Microlensing}

\author[Soton]{A.M. Newsam\thanksref{email}}
\author[Gla]{J.F.L. Simmons},
\author[Gla]{M.A. Hendry},
\author[Gla]{I.J. Coleman\thanksref{BAS}}
\address[Soton]{Department of Physics and Astronomy, University of Southampton,
 Southampton, UK}
\address[Gla]{Department of Physics and Astronomy, University of Glasgow,
 Glasgow, UK}

\thanks[email]{E-mail: amn@astro.soton.ac.uk}
\thanks[BAS]{Current address: British Antarctic Survey, Cambridge, CB3 0ET, UK}

\begin{abstract}
It has already been shown that microlensing can give rise to a
non-zero variable polarisation signal. Here we use realistic
simulations to demonstrate the additional information that can be
gained from polarimetric observations of lensing events.

\end{abstract}

\end{frontmatter}

\section{Introduction}
\label{intro}

It has been shown by \citeasnoun{Simmons+95b} (hereafter SNW95) that
microlensing of stars can produce a distinctive and significant
polarisation signature, even for lensed stars with no net unamplified
polarisation. In this contribution, we use the model developed in
SNW95 to simulate a series of realistic observations
of microlensing events. We use these simulations to examine how much
additional information about the parameters of the lensing event can
be determined from polarisation measurements.

%
\section{Microlensing induced polarisation of unpolarised stars}
\label{sec:model}

Normally, the light received from stars is unpolarised unless the star
has some suitable asymmetry (eg an asymmetric wind or radial
distortion). However, the emission from the limb of the star can be
expected to be quite strongly polarised. Therefore, during a
microlensing event, different parts of the limb may be amplified by
different amounts, breaking the symmetry of the star and giving rise
to an observable net polarisation.

This effect will be most significant if the stellar radius is
comparable to or larger than the einstein radius of the lens projected
onto the plane of the lensed star. The observed polarisation will
reach a maximum as the line of sight to the lens crosses the limb of
the star. This gives rise to the distinctive polarisation profiles of
lensing events given in SNW95, with a polarisation peak coincident with
the amplification peak if the line of sight to the lens remains
outside the stellar disk, but a double-peaked polarisation profile if
the lens appears to cross the disk of the star (``transit'' events).

Clearly, when studying an actual lensing event, one would use a model
appropriate to the observed stellar type. Also, one would hope to be
able to obtain information about the lensed star itself, as well as
the parameters of the lensing event
\citeaffixed{V-G98,Coleman+97}{e.g.}. However, here we will only 
consider the lensing parameters themselves and so we consider a simple
``generic'' stellar atmosphere model. The full details of the models
we use for the stellar atmosphere and microlensing event are given in
SNW95.
%

%
\begin{figure}
\begin{center}
\includegraphics*[width=0.4\hsize]{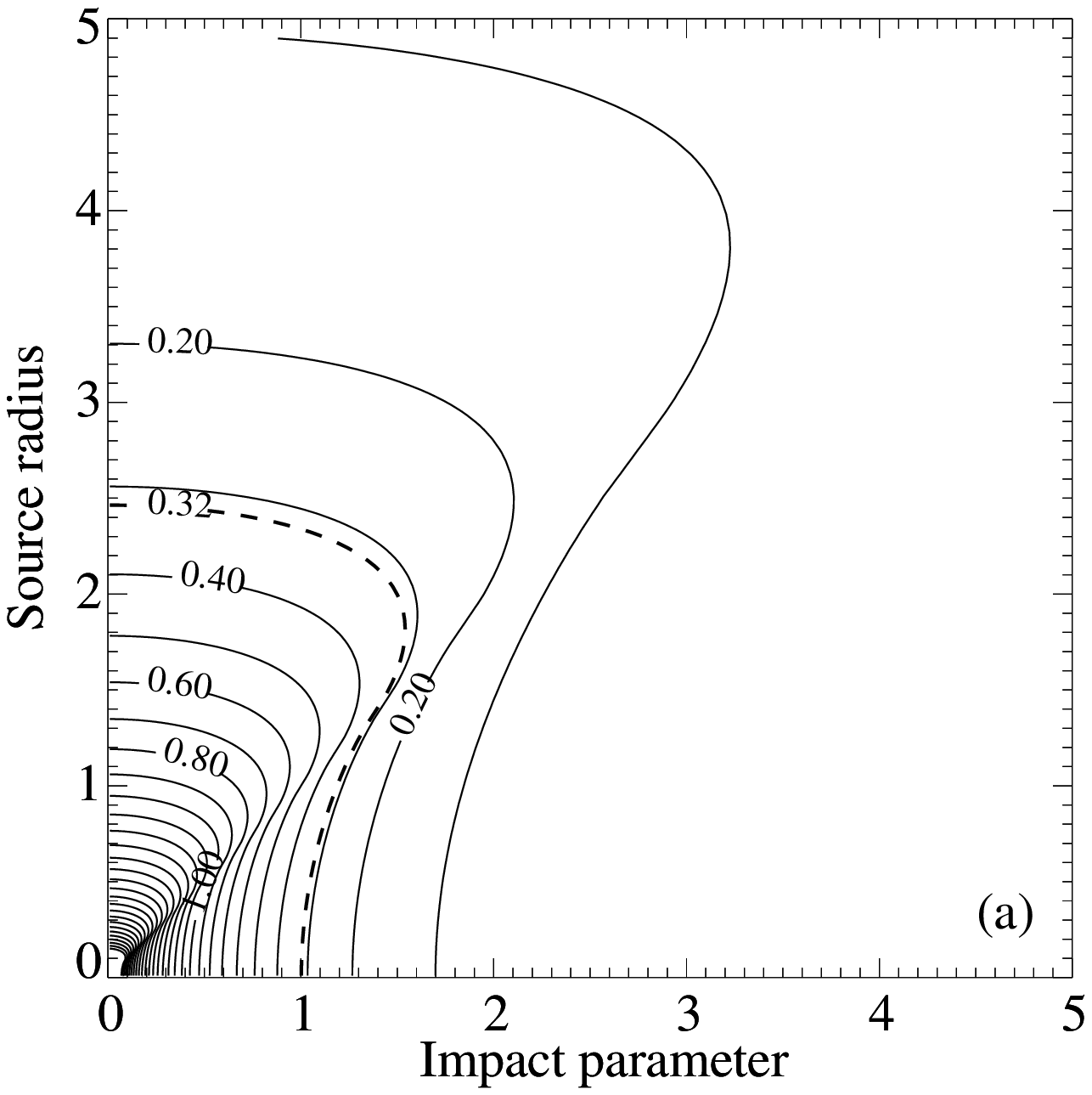}
\hspace{0.13\hsize}
\includegraphics*[width=0.4\hsize]{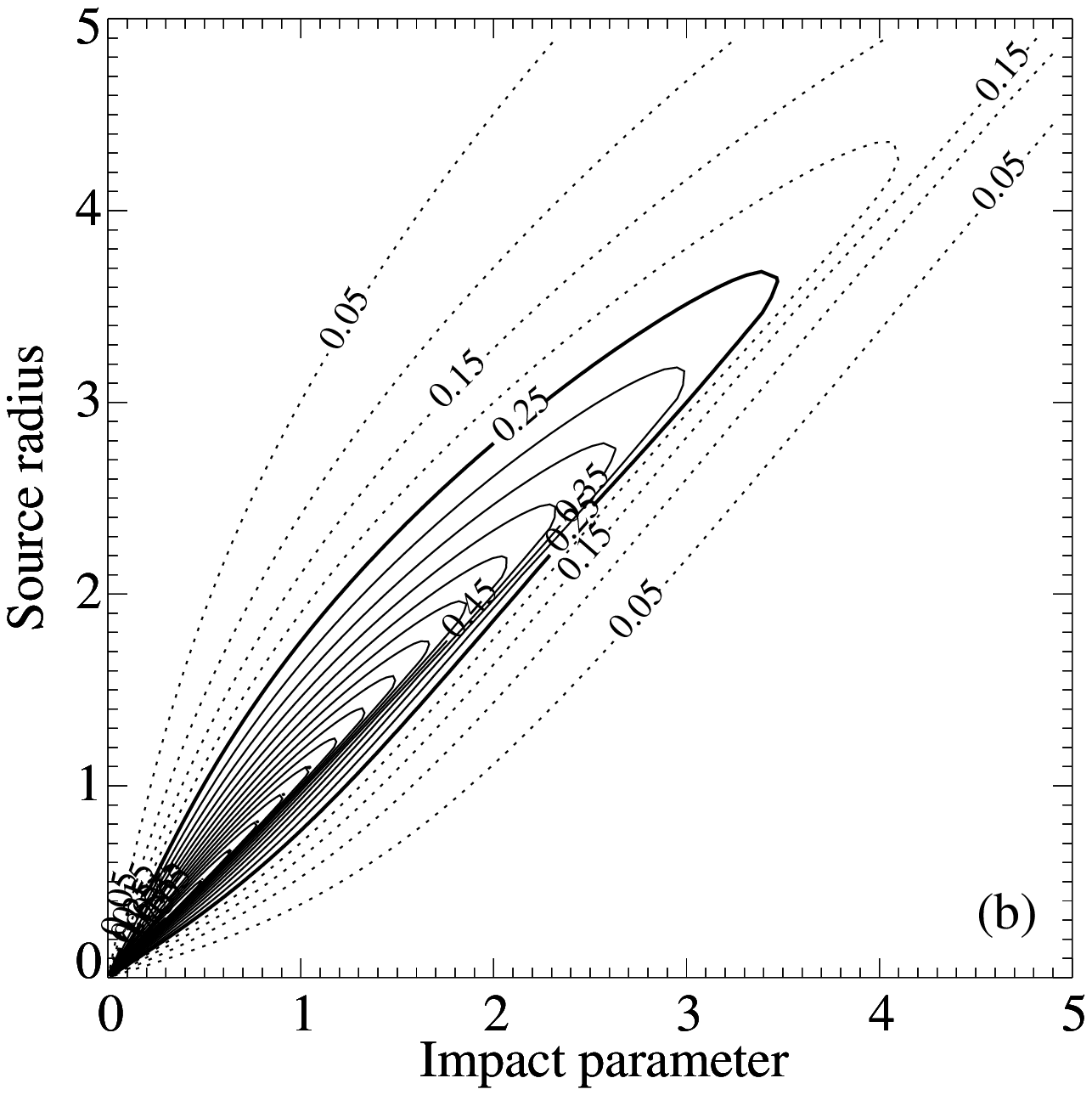}
\end{center}
\caption{Flux amplification and polarisation for microlensing of a simple 
stellar model. Both figures show the results for a range of source
radii and impact parameters $d_0$, both in units of the angular
einstein radius. Figure (a) shows the photometric amplification with
contours spaced by 0.1 magnitudes. The dashed contour is for an
amplification of 0.32 mag.~--~that expected in the point-source
approximation with $d_0=1$. Figure (b) shows the percentage
polarisation observed. Contours are at intervals of 0.05\% with dashed
contours below 0.25\%.}
\label{fig:modres}
\end{figure}
Figure~\ref{fig:modres} shows the results of applying a lensing
amplification function to this model for a range of sounce/lens
configurations. As shown in SNW95, when the source radius $R$ is very
much less than the projected distance between the star and the lens
(the impact parameter $d_0$), the star is effectively a point source
and there is no measurable polarisation. However, when $R \lesssim
d_0$, significant polarisation can occur, and the flux amplification
is modified from that expected for the point-source approximation
\citeaffixed{Gould94,WittMao94}{see also}.

\section{Simulations of realistic observations}
\label{sec:simul}

\begin{figure}
\begin{center}
\includegraphics*[width=0.42\hsize,height=0.35\hsize]{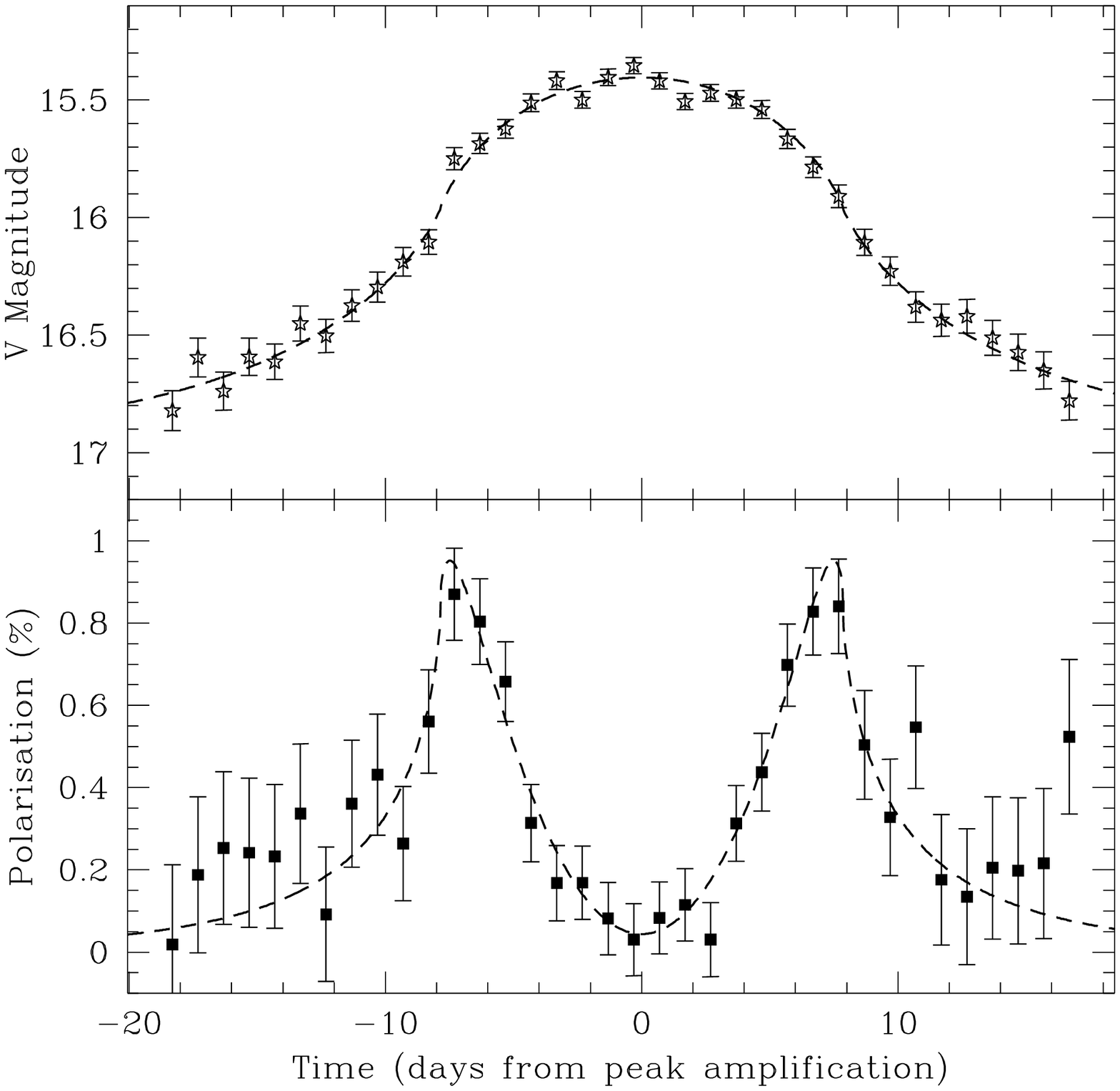}
\hspace{0.05\hsize}
\includegraphics*[width=0.42\hsize,height=0.35\hsize]{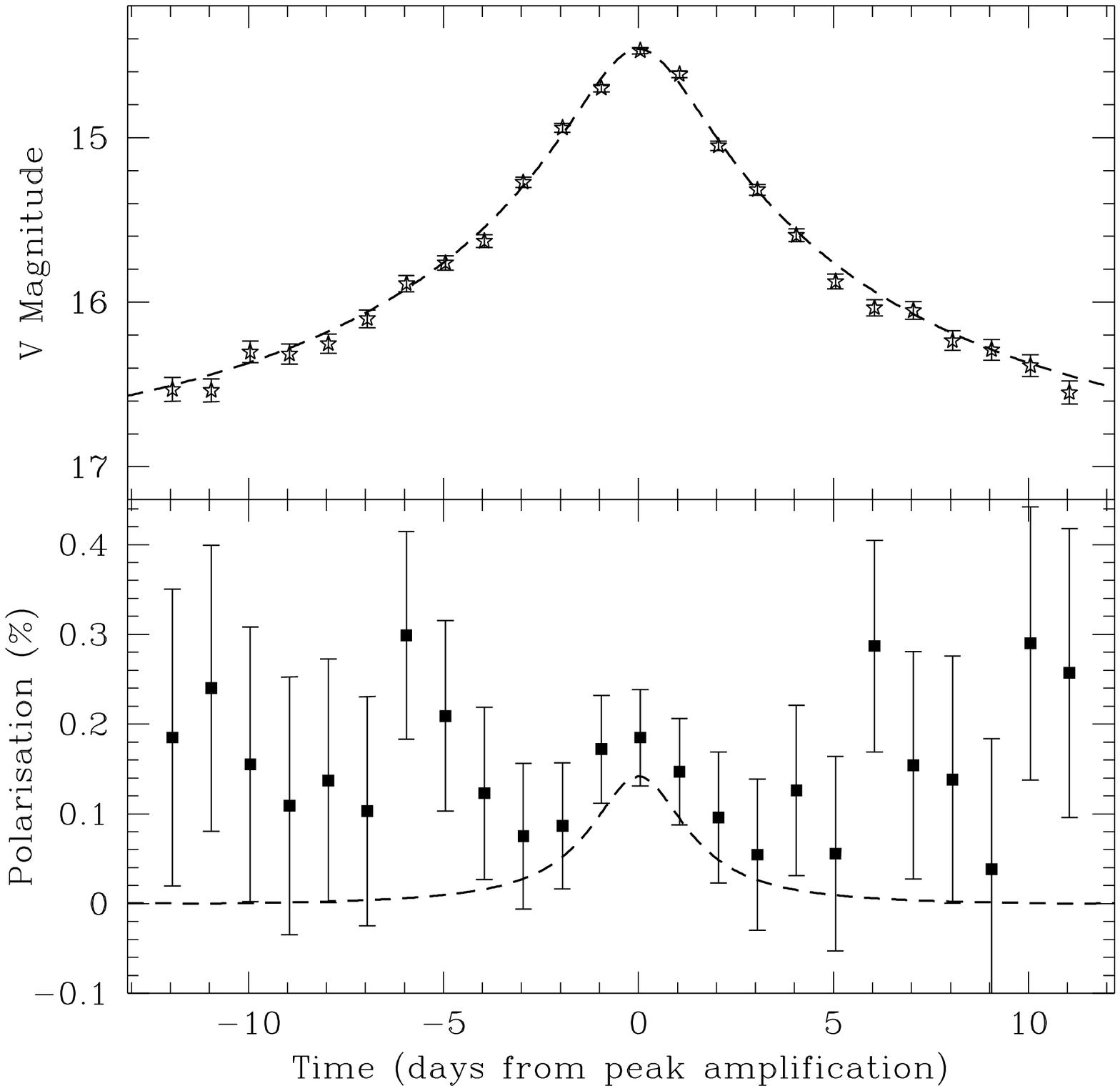}
\includegraphics*[width=0.42\hsize]{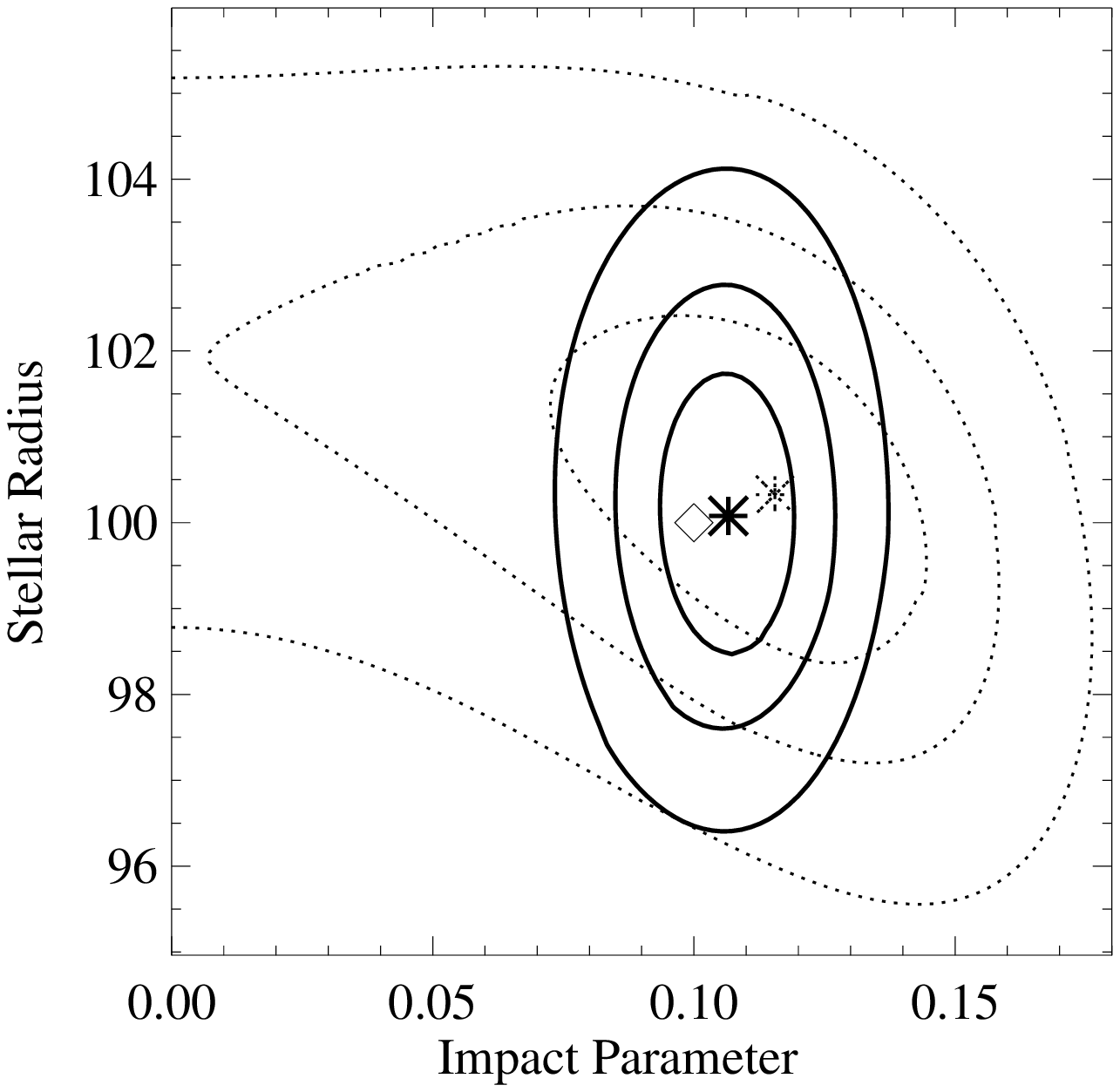}
\hspace{0.05\hsize}
\includegraphics*[width=0.42\hsize]{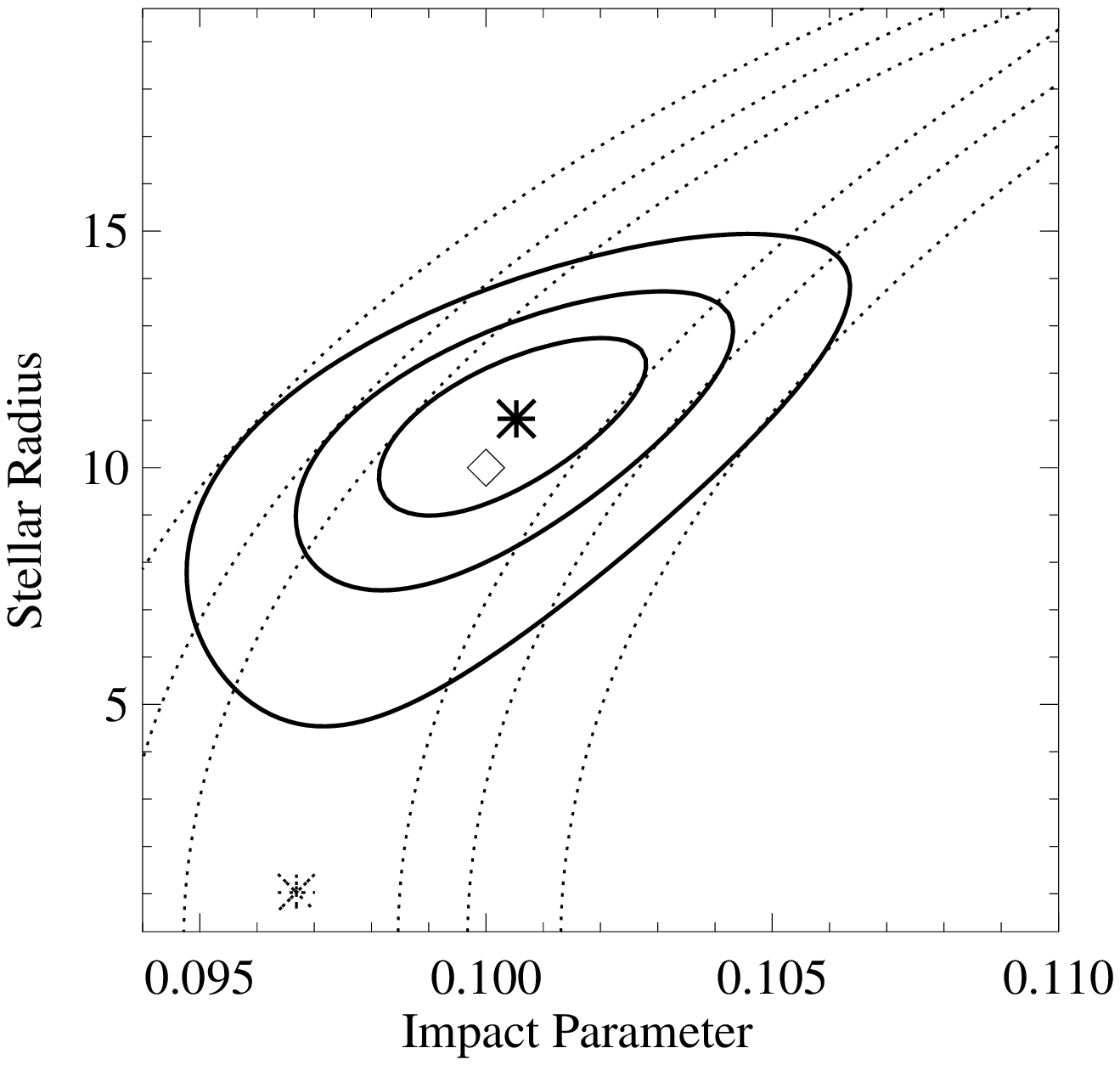}
\end{center}
\caption{Simulated observations and fits.
The upper  panels show the model photometric and polarimetric
profiles and the simulated observations. The lower panels show
the $\chi^2$ surfaces of two fits to the data. The thicker, solid
contours and star show the $\chi^2$ surface and best fit results for a
fit to both the flux and polarimetric data simultaneously. The dashed
lines and star show the result of a fit to just the photometric
data. The diamond shows the true solution. Contours are 68\%, 95\% and
99.7\% confidence regions. The left-hand panels are for an $R=100 {\rm
R}_\odot$ star, the right-hand panels for $R=10 {\rm R}_\odot$. It should be
noted that the fit actually gives the stellar radius in units of the
angular einstein radius, but this has been converted to Solar radii
for clarity.}
\label{fig:r100+r10}
\end{figure}

Using this model, we have simulated photometric and polarimetric
observations of microlensing events with realistic sampling and
errors. Two such simulations are given here (selected at random from
our ``catalogue''). Both simulations are for a V=17 star in the LMC
lensed by a 0.01 M$_\odot$ lens at half the distance to the LMC with a
velocity on the plane of the sky of 50 km/s. The impact parameter in
both cases is 0.1 angular einstein radii. Observations are taken once
``nightly''. Photometric errors are based on
\citeasnoun{Udalski+94} and polarimetric errors are calculated for a
2-hour exposure in V on a 1.2m telescope. The first simulation is for
an $R=100 {\rm R}_\odot$ source (figure~\ref{fig:r100+r10}, left - a
transit) and the second for $R=10 {\rm R}_\odot$
(figure~\ref{fig:r100+r10}, right). The figure also shows the results
of $\chi^2$ fits to the simulated data for two interesting parameters
(the stellar radius $R$ and the impact parameter $d_0$).
In both cases, the inclusion of polarimetric data provides a strong
constraint on the fit, In particular, in the 10R$_\odot$ case, where
the polarisation measurements are only just significant at the maximum
amplification, there is still useful information for determining the
stellar radius.

\section{Conclusions}
\label{sec:conclusion}

We have shown that microlensing can produce a significant, distinctive
polarisation signature and that observations of these signatures can
provide considerable additional information about the parameters of a
lensing event. In particular, some of the degeneracy between the
various parameters of the lensing event present in purely photometric
observations can be broken.  The realistic simulated observations
presented here demonstrate that such observations can be obtained with
modest resources. In a future paper (Newsam et al, in preparation) we
will present a thorough exploration of the lensing parameter-space
with more realistic stellar atmosphere models both to determine when
polarisation measurements would be most useful, and to examine the use
of such measurements to study the atmosphere of the {\em lensed\/}
star \citeaffixed{Coleman+97}{see}.

\end{document}